\documentclass[%
nofootinbib,
superscriptaddress,
 preprint,
showpacs,preprintnumbers,
 amsmath,amssymb,
 aps,
pra,
 longbibliography,
 lengthcheck,%
]{revtex4-1}

\usepackage{quantum_friction_2}

\let\oldsection\section
\renewcommand{\section}[1]{\emph{#1}.}

\newcommand*{\appref}[1]{
\def\temp{#1}\ifx\temp\empty{\leavevmode\unskip supplemental material}\ignorespaces
\else
    \leavevmode\unskip Appendix~\ref{#1}\ignorespaces
\fi
}

\begin{document}

\title{No thermalization without correlations}

\author{Dmitry V. Zhdanov}
\email{dm.zhdanov@gmail.com}
\affiliation{Northwestern University, Evanston, Illinois 60208, USA}
\author{Denys I. Bondar}
\affiliation{Princeton University, Princeton, New Jersey 08544, USA}
\author{Tamar Seideman}
\affiliation{Northwestern University, Evanston, Illinois 60208, USA}
\begin{abstract}
%
%
%
%
The proof of the long-standing conjecture is presented that Markovian quantum master equations are at odds with quantum thermodynamics under conventional assumptions of fluctuation-dissipation theorems (implying a translation invariant dissipation). Specifically, except for identified systems, persistent system-bath correlations of at least one kind, spatial or temporal, are obligatory for thermalization.
A systematic procedure is proposed to construct translation invariant bath models producing steady states that well-approximate thermal states.
A quantum optical scheme for the laboratory assessment of the developed procedure is outlined.
%
%
%
\end{abstract}
\maketitle
\section{Introduction}
A stochastic interaction of a quantum system with a bath 
brings up the term $\ehat\ffrnforce$ in the relations for time-dependent expectation values of system momenta $\hat\pp{=}\{\hat p_1,\ldots,\hat p_N\}$ and positions $\hat\xx{=}\{\hat x_1,\ldots,\hat x_N\}$:%
\begin{subequations}\label{ODM-eqs}
\begin{gather}
\label{ODM-dp/dt}
\tder{}{t}\midop{\hat p_n}{=}{-}\midop{\tpder{}{\hat x_n}U(\hat \xx)}{+}\midop{\ehat\frnforce_n
},\\
\label{ODM-dx/dt}
\tder{}{t}\midop{\hat x_n}{=}\tfrac{1}{m_n}\midop{\hat p_n},
\end{gather}
\end{subequations}
where $U(\hat \xx)$ is a potential energy operator and $m_k$ are effective masses. In this Letter, we study the case where $\ehat\ffrnforce{=}\ehat\ffrnforce(\hat{\pp})$ is position-independent. In this form, Eqs.~\eqref{ODM-eqs} apply to many quantum phenomena including the translational motion of an excited atom in vacuum \cite{2017-Sonnleitner}, Brownian motion in a dilute background gas \cite{2003-Accardi}, light-driven processes in semiconductor, nanoplasmonic and optomechanical systems \cite{BOOK-Auffeves,BOOK-Milburn,2015-Barchielli}, superconducting currents \cite{1996-OConnell}, quantum ratchets \cite{2002-Reimann}, energy transport in low-dimensional systems \cite{2008-Dhar}, dynamics of chemical reactions \cite{2000-Kuhn}, two-dimensional vibrational spectroscopy and NMR signals \cite{2000-Steffen,2006-Tanimura} as well as more exotic entirely quantum dissipative effects \cite{2010-Rezek,1999-Kardar}. 

The term $\ehat\ffrnforce(\hat{\pp})$ in Eqs.~\eqref{ODM-eqs} admits  a simple classical interpretation as friction acting on effective particles moving in a potential $U(\xx)$.
Such classical dynamics are described by the familiar Langevin, Drude and Fokker-Plank models when the system-bath interactions are treated as 
\begin{enumerate*}[label=(\roman*)]
\item \label{A_i}
memoryless (Markovian) and
\item \label{A_ii}
translation invariant (position-independent).
\end{enumerate*}
However, we will show that these two assumptions are at odds with quantum thermodynamics.
Specifically, we will prove a long-standing no-go conjecture that completely positive\footnote{Positivity of quantum evolution guarantees satisfaction of the Heisenberg uncertainty principle at all times. It was shown that the requirements for positivity and complete positivity coincide for some quantum systems including a harmonic oscillator \cite{1997-Kohen}.} Markovian translation-invariant quantum dynamics obeying Eqs.~\eqref{ODM-eqs} cannot thermalize. 

The no-go conjecture was demonstrated by Lindblad as early as in 1976 \cite{1976-Lindblad-a} for a quantum harmonic oscillator with a Gaussian damping%
\footnote{The Gaussian damping corresponds to $\Lrel{=}\Lbd_{\set{\mu}\hat\xx{+}\set{\eta}\hat\pp}$ ($\set{\mu},\set{\eta}\in \mathbb{C}^N$) in Eq.~\eqref{Quantum_Liouville_equation_nD(a)} and can be cast to form \eqref{theorem:trans_inv_Lbd}, as shown in \appref{@APP:theo:trans_inv_Lbd}). The original paper \cite{1976-Lindblad-a} deals with one-dimensional case. The multidimensional extension can be found e.g. in \cite{1985-Dodonov}.}.
Subsequently his particular result was extended to a general quantum system under the weight of mounting numerical evidence, however without proof. The no-go conjecture is de-facto incorporated in all popular models such as the Redfield theory \cite{1957-Redfield}, the Gaussian phase space ansatz of Yan and Mukamel \cite{1988-Yi_Jing_Yan}, the master equations of Agarwal \cite{1971-Agarwal}, Caldeira-Leggett \cite{1983-Caldeira}, Hu-Paz-Zhang \cite{2005-Ford}, and Louisell/Lax \cite{1965-Louisell}, and the semigroup theory of Lindblad \cite{1976-Lindblad} along with specialized extensions in different areas of physics and chemistry. These models break either one of assumptions \ref{A_i} and \ref{A_ii} or the complete positivity of quantum evolution (see \cite{1997-Kohen,2000-Yan,2016-Bondar} for detailed reviews, note errata \cite{2016-Bondar-a}). This circumstance is a persistent source of controversies (see e.g.~the discussions \cite{1998-Wiseman,2001-OConnell,2001-Vacchini} of original works \cite{1997-Gao,2000-Vacchini}).
The matters were further complicated by the discovery that the free Brownian motion $U(\hat\xx){=}0$ circumvents the conjecture \cite{2002-Vacchini} (we will identify the full scope of possible exceptions below).

The no-go result challenges studies of the long-time dynamics of open systems. On the one hand, model's thermodynamic consistency is undermined by assumptions \ref{A_i} and \ref{A_ii}. On other hand, the same assumptions open opportunities to simulate large systems that are otherwise beyond the reach. Specifically, the abandonment of Markovianity 
entails a substantial overhead to store and process the evolution history. The value of assumption \ref{A_ii} can be clarified by the following example. Consider the re-thermalization of a harmonic oscillator coupled to a bath (represented by a collection of harmonic oscillators) after displacement from equilibrium by, e.g., an added external field, a varied system-bath coupling, or interactions between parts of a compound system. 
To account for such a displacement without assumption \ref{A_ii}, one needs to self-consistently identify the equilibrium position for each bath oscillator, re-thermalize the bath and modify the system-bath couplings accordingly. In practice, this procedure is intractable 
 without gross approximations that lead to either numerical instabilities or physical inaccuracies. Choosing among a polaron-transformation-based method, Redfield, and F\"orster (hopping) models of quantum transfer epitomizes this dilemma \cite{2013-Jang}. 

\begin{figure}[tbp]
\centering\includegraphics[width=0.99\columnwidth]
{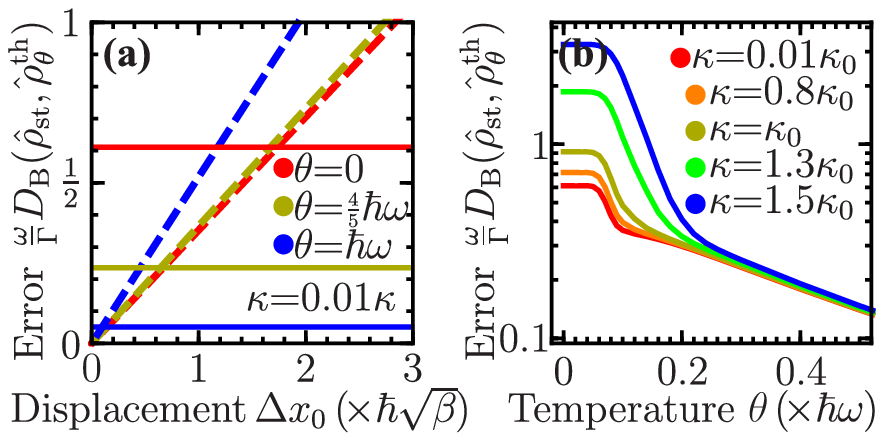}
\caption{
The errors (expressed in the terms of Bures distance $D_{\idx{B}}$ between the  thermal state $\rhoth{\theta}$ and its approximation $\rhost$) in modeling thermal states of a 1D quantum harmonic oscillator in the displaced equilibrium configurations (due to a change $U(\hat x){\to}U(\hat x{-}\Delta x_0)$ in the potential energy) using the conventional
quantum optical master equation (dashed lines) and the proposed translation-invariant dissipation model defined by Eqs.~\eqref{Quantum_Liouville_equation_nD},\eqref{theorem:trans_inv_Lbd} and \eqref{_optimal_f_k} (solid lines). 
(a) The error dependence on displacement $\Delta x_0$ for several temperatures $\theta$. (b) The error dependence on temperature $\theta$ for different values of $\kappa$ (in units of $\kappa_0{=}\hbar^{-1}\beta^{-\frac12}$).\label{@FIG.01b'}}
\end{figure}

Remarkably, assumption \ref{A_ii} enables to model the displaced state equilibrium by simply adjusting the potential energy $\hat U$. Fig.~\ref{@FIG.01b'}a shows that without this assumption the potential adjustment yields steady state $\rhost$ significantly different from the canonical equilibrium $\rhoth{\theta}{\propto}e^{{-}\frac{\hat H}{\theta}}$, where $\theta{=}k_{\idx{B}}T$ and $\hat H$ is system Hamiltonian. 


Motivated by these arguments, we propose in this Letter a general recipe to construct approximately thermalizable bath models under assumptions \ref{A_i} and \ref{A_ii}. Fig.~\ref{@FIG.01b'} illustrates this recipe in application to the above example. 
The resulting mismatch between $\rhost$ and $\rhoth{\theta}$ is small, especially at high temperatures and in the weak system-bath coupling limit. (The calculations details will be explained below.)

It will be shown elsewhere that the proposed recipe is capable of accurately accounting for electronic and spin degrees of freedom. We found it helpful in reservoir engineering and optimal control problems. Moreover, the resulting bath models are realizable in the laboratory and can be used for coupling atoms and molecules nonreciprocally \cite{2016-Zhdanov}. However, the scope of our recipe is limited by the applicability of assumptions \ref{A_i} and \ref{A_ii} and, therefore, cannot encompass strongly correlated systems (as in the case of  Anderson localization \cite{2015-Nandkishore}).  

\section{The key results}
Starting by formalizing the problem, we write the general master equation that accounts for memoryless system-bath interactions and ensures positivity of the system density matrix $\hat\rho$ at all times \cite{1976-Lindblad}:%
\begin{subequations}
\label{Quantum_Liouville_equation_nD}
\begin{gather}\label{Quantum_Liouville_equation_nD(a)}
\tpder{}{t}\hat\rho{=}{\cal L}[\hat\rho],~{\cal L}{=}\Lvn{+}\Lrel,\\
%
\Lvn[\odot]{=}\tfrac{i}{\hbar}[\odot,\hat H],~\hat H{=}H(\hat\pp,\hat\xx){=}\textstyle\sum_{n=1}^N\frac{\hat p_n^2}{2 m_n}{+}U(\hat\xx),\label{Quantum_Hamiltonian_nD}\\[-4pt]
\Lrel{=}\sum_{k{=}1}^K\Lbd_{\hat L_k},
\label{_Lindbladian_definition}
~~
\Lbd_{\hat L}[\hat\rho]{\defeq}\hat L{\hat\rho}\hat L^{\dagger}{-}\tfrac12(\hat L^{\dagger}\hat L{\hat\rho}{+}{\hat\rho}L^{\dagger}\hat L),
\end{gather}
\end{subequations}
where $\odot$ is the substitution symbol defined, e.g., in \cite{2000-Grishanin}. The superoperator $\cal{L}_{\idx{rel}}$ accounts for system-bath couplings responsible for the friction term $\ehat\ffrnforce$ in Eq.~\eqref{ODM-dp/dt} and depends on a set of generally non-Hermitian operators $\hat L_k$.
Based on theorems by A. Holevo \cite{1995-Holevo,1996-Holevo}, B. Vacchini \cite{2005-Petruccione,2005-Vacchini,2009-Vacchini} has identified the following criterion of translational invariance for the $\cal{L}_{\idx{rel}}$:
\begin{lemma}[The justification is in \appref{@APP:theo:trans_inv_Lbd}]\label{@theo:trans_inv_Lbd}
Any translationally invariant superoperator $\Lrel$ of the Lindblad form \eqref{_Lindbladian_definition} can be represented as%
\begin{subequations}\label{theorem:trans_inv_Lbd}
\begin{gather}
\displaybreak[0]\Lrel{=}\textstyle\sum_k\Lbd_{\hat A_k}+\Laux \mbox{ with } \displaybreak[0]\\
\hat A_k{\defeq}e^{{-}i\kkappa_k\hat\xx}\tilde f_k(\hat\pp),~~\Laux{=}{-}i[\kkappaaux\hat\xx{+}\faux(\hat \pp),\odot].
\end{gather}
\end{subequations}
where $\kkappa_k$ and $\kkappaaux$ are $N$-dimensional real vectors, $\tilde f_k$ are complex-valued functions and $\faux$ is real-valued
\footnote
{
The Gaussian dissipators $\Lbd_{\kkappaG_k\hat\xx{+}\ffG_k(\hat\pp)}$ $(\kkappaG_k{\in}\mathbb{R}^N)$ can be reduced to the form 
Eq.~\eqref{theorem:trans_inv_Lbd} as a limiting case $\kkappa_k{\to}0$, as shown in  \appref{@APP:theo:trans_inv_Lbd}. The 
generalized unitary drift term $\Laux$ accounts for ambiguity of the separation of the quantum Liouvillian ${\cal L}$ in Eq.~\eqref{Quantum_Liouville_equation_nD(a)} into Hamiltonian and relaxation parts.
}
.The converse holds as well.
\end{lemma}

The primary findings of this work are summarized in the following two no-go theorems.
\begin{theorem} \label{@theo:no_go(T=0)}
Let $\ket{\Psi_0}$ be the ground state (or any other eigenstate of $\hat H$), such that $\matel{\Psi_0}{\hat\pp}{\Psi_0}{=}0$, and which momentum-space wavefunction $\Psi_{0}(\pp){=}\scpr{\pp}{\Psi_0}$ is nonzero almost everywhere, except for some isolated points. Then, no translationally invariant Markovian process of form \eqref{Quantum_Liouville_equation_nD} and \eqref{theorem:trans_inv_Lbd} can steer the system to $\ket{\Psi_0}$.
\end{theorem}

The idea of the proof, whose details are given in \appref{@APP:theo:no_go(T=0)}, is to show that the state $\hat\rho_{0}{=}\proj{\Psi_{0}}$ can be the fixed point of superoperator $e^{t\cal L}$ only if $\Lrel{\equiv}0$. First, note that the linearity and translation invariance of the dissipator \eqref{theorem:trans_inv_Lbd} imply that $\Lrel[\int g(\xx')e^{{-}\frac{i}{\hbar}\xx'\hat\pp}\hat\rho_{0}e^{\frac{i}{\hbar}\xx'\hat\pp}d^N\xx']{=}0$ for any function $g(\xx')$. This equation can be equivalently rewritten as
\begin{gather}\label{ngt1_sample_eq}
\Lrel[\Psi_{0}(\hat\pp)g(\hat\xx)\Psi_{0}(\hat\pp)^{\dagger}]{=}0
\end{gather}
using the identities $e^{{-}\frac{i}{\hbar}\xx'\hat\pp}\ket{\Psi_{0}}{=}\sqrt{2\pi\hbar}\Psi_{0}(\hat\pp)\ket{\xx'}$ and $\int g(\xx')\proj{\xx'}d^N\xx'{=}g(\hat\xx)$, where $\ket{\xx'}$ is the eigenstate of position operator: $\hat x_k\ket{\xx'}{=}x'_k\ket{\xx'}$. Let us choose $g(\xx){=}e^{-i\llambda\xx}$, where  $\llambda$ is an arbitrary real vector, and move to the right the $\hat\xx$-dependent terms in the lhs of Eq.~\eqref{ngt1_sample_eq} using the commutation relations $e^{-i\tilde\llambda\hat\xx}\hat\pp = (\hat\pp{+}\hbar\tilde\lambda)e^{-i\tilde\llambda\hat\xx}$ with $\tilde\llambda{=}\llambda,\pm\kkappa_k$. This rearrangement brings Eq.~\eqref{ngt1_sample_eq} to the form $\tilde G_{\llambda}(\hat\pp)e^{-i \llambda\hat{\xx}}{=}0$ (note that all the operators of form $e^{\pm i\tilde\kkappa_k\hat\xx}$ expectedly cancel out owing to translation invariance of $\Lrel$). The last equality can be satisfied only if the function $\tilde G_{\llambda}(\pp)$ vanishes identically for all $\pp$ and $\llambda$. 
However,  careful inspection of \appref{@APP:theo:no_go(T=0)}~shows that the latter happens only if $\Lrel{=}0$.


The statement of the \ref{@theo:no_go(T=0)}-st no-go theorem can be strengthened for a special class of quantum systems. Let $\BFn(\pp,\llambda)$ be the Blokhintsev function \cite{1940-Blokhintzev}, which is related to Wigner quasiprobability distribution $W(\pp,\xx)$ as
\begin{gather}\label{_Blokhintsev_function}
\BFn(\pp,\llambda){=}\textstyle\int_{{-}\infty}^{\infty}\ldots\textstyle\int_{{-}\infty}^{\infty} e^{i \llambda\xx} W(\pp,\xx)\diff^N\xx.
\end{gather}

\begin{theorem}[]\label{@theo:no_go(T>0)} 
Suppose that the Blokhintsev function $\BFn_{\theta}(\pp,\llambda)$ of the thermal state $ \rhoth{\theta}{\propto}e^{-\frac{\hat H}{\theta}}$ characterized by temperature $k_{\idx{B}}T{=}\theta$ is such that%
\begin{subequations}\label{_B(p,lambda)-features}
\begin{gather}\label{_B(p,lambda)-features-a}
\forall\pp,\llambda: \BFn_{\theta}(\pp,\llambda){>}0,~~\BFn_{\theta}(\pp,{-}\llambda){=}\BFn_{\theta}(\pp,\llambda),\\
%
\forall\pp{\ne}\zzero,\llambda{\ne}\zzero: \BFn_{\theta}(\pp,\llambda){<}\BFn_{\theta}(\zzero,\zzero).\label{_B(p,lambda)-features-b}
\end{gather}
\end{subequations}
Then, no translationally invariant Markovian process \eqref{Quantum_Liouville_equation_nD} and \eqref{theorem:trans_inv_Lbd} can asymptotically steer the system to $\rhoth{\theta}$.
\end{theorem}

The proof of this theorem is given in \appref{@APP:theo:no_go(T>0)}~and generally follows the same logic as the outlined proof of the \ref{@theo:no_go(T=0)}-st no-go theorem.
Using Eq.~\eqref{_Blokhintsev_function} and the familiar formula for the thermal state Wigner function \cite{1949-Bartlett}, it is easy to check that the criteria \eqref{_B(p,lambda)-features} are satisfied for any $\theta$ in the case of a quadratic potential $U$. This means that the Lindblad's original conclusion on inability to thermalize the damped harmonic oscillator using the Gaussian friction term $\Lrel{=}\Lbd_{\set{\mu}\hat\xx{+}\set{\eta}\hat\pp}$ is equally valid for all Markovian translationally invariant dissipators. 

\begin{corollary}\label{@cor:no_go_qho(T>0)}
No translationally invariant Markovian process of form \eqref{Quantum_Liouville_equation_nD} and \eqref{theorem:trans_inv_Lbd} can steer the quantum harmonic oscillator into a thermal state of form $ \rhoth{\theta}{\propto}e^{{-}\frac{\hat H}{\theta}}$.
\end{corollary}

\section{Practical implications of the no-go theorems} In classical thermodynamics, the bath is understood as a constant-temperature heat tank ``unaware'' of a system of interest. However, the no-go theorems indicate that system-bath correlations of at least one kind -- spatial or temporal -- become obligatory for thermalization once quantum mechanical effects are taken into account. These correlations break the bath translation invariance or Markovianity assumptions, respectively. 


Nevertheless, in the view of computational advantages outlined above, it is desirable to incorporate these assumptions into the master equations \eqref{Quantum_Liouville_equation_nD} and \eqref{theorem:trans_inv_Lbd}. Now we are going to introduce the recipe to construct such models with a minimal error in the thermal state.
In order to proceed, note that in the limit $(\hbar\kkappa_k)^2{\ll}\midop{\hat\pp^2}$ Eqs.~\eqref{Quantum_Liouville_equation_nD} and \eqref{theorem:trans_inv_Lbd} reduce to the familiar Fokker-Planck equation
\begin{gather}
\tpder{}{t}{\varpi(\pp)}{\stackrel{}{\simeq}}\Tr[\delta(\pp{-}\hat\pp)\Lvn[\hat\rho]]{+}\notag\\
\sum_{n,l}\pder{^2D_{n,l}(\pp)\varpi(\pp)}{p_n\partial p_l}{-}\sum_{n}\pder{\frnforce_n(\pp)\varpi(\pp)}{p_n}\label{_Fokker_Planck_equation}
\end{gather}
for the momentum probability distribution $\varpi(\pp){=}\Tr[\delta(\pp{-}\hat\pp)\hat\rho]$. The friction forces $\ffrnforce$ in Eq.~\eqref{_Fokker_Planck_equation} as well as Eq.~\eqref{ODM-dp/dt} have the form 
\begin{gather}\label{_Bondarian_gen}
{\ffrnforce}(\hat\pp){=}{-}\textstyle\sum_k\hbar\kkappa_{k}|\fc_k(\hat\pp)|^2,
\end{gather}
whereas the momentum-dependent diffusion operator is
\begin{gather}\label{_Diffusion_operator}
D_{n,l}(\hat\pp){=}\textstyle\tfrac{\hbar^2}2\textstyle\sum_k|\fc_k(\hat\pp)|^2\kappa_{k,n}\kappa_{k,l}.
\end{gather}

Equations~\eqref{_Bondarian_gen} and \eqref{_Diffusion_operator} can be satisfied by different sets of $\kkappa_k$ and $\tilde f_k(p)$. We will exploit this non-uniqueness to reduce the system-bath correlation errors. Our strategy is reminiscent to the familiar way of making density functional calculations practical via error cancellation in approximated exchange-correlation functionals. We shall demonstrate the generic procedure 
by considering a one-dimensional oscillator with the Hamiltonian $\hat H{=}\tfrac m2\hat p^2{+}\frac{m\omega^2}2\hat x^2$ (here the dimension subscript $n$ is omitted for brevity).
Corollary~\ref{@cor:no_go_qho(T>0)} implies that $\Lrel[\rhoth{\theta}]{\ne}0$ and $\rhost{\ne}\rhoth{\theta}$ for any $\theta$, where $\rhost{=}\left.\hat\rho\right|_{t\to\infty}$ is the actual fixed point of the evolution operator $e^{t{\cal L}}$. However, the net discrepancies can be reduced by imposing the following thermal population conserving constraint:
\begin{gather}\label{_energy_distribution_conservation}
\left.\tder{}{t}\midop{e^{-\alpha\hat H}}_{\theta}\right|_{t{=}0}{=}0;~\left|\tder{^2}{t^2}\midop{e^{-\alpha\hat H}}_{\theta}\right|_{t{=}0}{\to}\min \mbox{ for all }\alpha,
\end{gather}
where $\midop{\odot}_{\theta}(t){=}\Tr[\odot e^{t{\cal L}}[\rhoth{\theta}]]$. This constraint can be intuitively justified when the characteristic decay rates are much smaller than the typical transition frequencies, such that the dissipation can be treated perturbatively. Since the term $\Lrel[\rhoth{\theta}]$ generates only rapidly oscillating off-diagonal elements in the basis of $\hat H$, Eq.~\eqref{_energy_distribution_conservation} ensures that the first-order perturbation vanishes on average for the exact thermal state: $\lim_{t\to{\infty}}\tfrac{1}t\int_0^{t}e^{\tau\Lvn}\Lrel e^{(t-\tau)\Lvn}[\rhoth{\theta}]\diff\tau{=}0$. 

In the case of the driftless dissipation $\Laux{=}0$, Eq.~\eqref{_energy_distribution_conservation} is satisfied by the following functions $\tilde f_{k}(p)$ in Eq.~\eqref{theorem:trans_inv_Lbd}:
\begin{gather}\label{_optimal_f_k}
\tilde f_{k}(p){=}c_k e^{p \beta \hbar  \lambda_{k}},~~\lambda_k{=}\kappa_{k}\tanh(\tfrac{\hbar\omega}{4\theta}),
\end{gather} 
where $\beta{=}(m\hbar\omega)^{-1}$ and the constants $c_k$ should be chosen to satisfy Eq.~\eqref{_Bondarian_gen}. The corresponding dissipator \eqref{theorem:trans_inv_Lbd} reproduces the familiar microphysical model of quantum Brownian motion (see e.g. Eq.~(16) in Ref.~\cite{2002-Vacchini}) in the limit $\kkappa{\to}0$, $\omega{\to}0$. Furthermore, the resulting dynamics tends to decrease (increase) the average system energy $\midop{\hat H}_{\theta}$ if its initial temperature $\theta'$ is higher (lower) than $\theta$:
\begin{gather}\label{_stability_check}
\tder{}{t}{\midop{\hat H}_{\theta'}}\big|_{t{=}0}{=}\tfrac{c_k^2}{\omega}\tilde\gamma^{\idx{en}}_k(\theta',\theta)(\midop{\hat H}_{\theta}{-}\midop{\hat H}_{\theta'})\big|_{t{=}0},
\end{gather}
where $\tilde\gamma^{\idx{en}}_k(\theta',\theta){=}2\omega{\beta\hbar^2\kappa_k \lambda_k \exp\left({{\beta\hbar^2\lambda_k^2}{\coth(\frac{\hbar\omega}{2 \theta' })}}\right)}{>}0$.

\begin{figure}[tbp]
\centering\includegraphics[width=\columnwidth]
{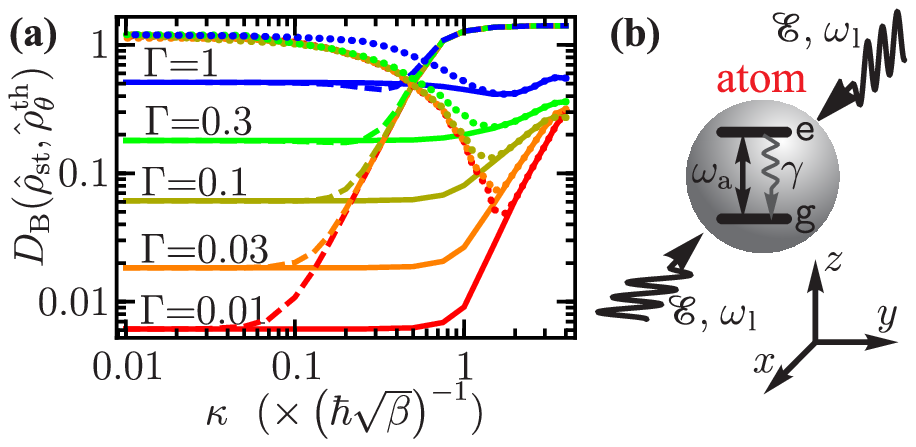}
\caption{(a) The accuracy of thermalization of the harmonic oscillator at $\theta{=}0$ by the dissipator $\Lrel{=}\Gamma\Bdn_{\kappa,\fciso}$ as function of $\kappa$ and $\Gamma$. The solid curves show the Bures distance $D_{\idx{B}}$ between   the  thermal state $\rhoth{\theta}$ and its approximation $\rhost$ for the case $\fciso(p)$ defined by Eq.~\eqref{_optimal_f_k} with $c{=}\omega/\sqrt{\tilde\gamma^{\idx{en}}(0,0)}$. The dotted curves represent the clipped versions \eqref{__clipping_rule} of $\fciso(p)$.
The dashed curves correspond to the case of functions $\fciso(p)$ approximated by Eq.~\eqref{_coherent_Doppler_f(p)} with parameters $\tilde c_{i}$ chosen such that $\left.\tder{^l}{p^l}(\fciso(p){-}\fcisoD(p))\right|_{p=0}{=}0$ for $l{=}0,1,2$.
(b) The Doppler cooling setup to test the model \eqref{Quantum_Liouville_equation_nD}, \eqref{theorem:trans_inv_Lbd} in the laboratory.\label{@FIG.01'}}
\end{figure}

Equation~\eqref{_stability_check} suggests that $\rhost$ is close to $\rhoth{\theta}$. This conclusion is supported by the simulations presented in Fig.~\ref{@FIG.01'}a for the isotropic dissipator $\Lrel{=}\Bdn_{\kappa,\fciso}$,
\begin{gather}\label{theorem:trans_inv_Lbd-isotropic}
\Bdn_{\kappa,\fciso}{\defeq}\Lbd_{\hat A^+}{+}\Lbd_{\hat A^-},~~\hat A^{\pm}{=}e^{{{\mp}}i\kappa\hat x}\fciso({\pm}\hat p).
\end{gather}
One can see that the high-quality thermalization is readily achieved by tuning the free parameters $c_k$ and $\kappa_k$ even in the strong dissipation regime. 

To understand the result \eqref{_optimal_f_k}, note that the terms $\Lbd_{\hat A_k}$ in Eq.~\eqref{theorem:trans_inv_Lbd} represent independent statistical forces $\midop{{-}\hbar\kkappa_{k}|\tilde f_k(\hat\pp)|^2}$ contributing to the net friction $\midop{\ehat\ffrnforce}$. 
In classical mechanics, such forces at $\theta{=}0$ steer the system to the state of rest by acting against the particles' momenta, hence 
\begin{gather}\label{__clipping_rule}
\tilde f_k(\hat\pp){=}0\mbox{ when }\pp\kkappa_k{<}0~~~\mbox{(classical mechanics)}. 
\end{gather} 
However, clipping the functions \eqref{_optimal_f_k} according to Eq.~\eqref{__clipping_rule} introduces significant errors, as displayed by dotted curves in Fig.~\ref{@FIG.01'}a. Thus, the ``endothermic'' tails of $\tilde f_k(\hat\pp)$ at $\pp\kkappa_k{>}0$ break the thermalization in the classical case, but reduce errors in the quantum mechanical treatment. To clarify this counterintuitive observation, note that the 
 physical requirement $\der{}{t}\midop{\hat O}_{\theta}{=}0$ for any observable $\hat O$ in the thermodynamic equilibrium $\rhost{=}\rhoth{\theta}$ is violated by the master equations \eqref{Quantum_Liouville_equation_nD} and \eqref{theorem:trans_inv_Lbd} due to the no-go theorems, i.e., 
\begin{gather}\label{_2-nd_moments_x^2}
\tder{}{t}{\midop{\hat x_n^2}_{\theta}}\bigg|_{t{=}0}{=}\hbar^2\sum_k\midop{\bigl|\tpder{}{\hat p_n}{
\fc_k(\hat\pp)
}\bigr|^2}_{\theta}\bigg|_{t{=}0}{>}0
\end{gather}
in the driftless case $\Laux{=}0$. The inequality~\eqref{_2-nd_moments_x^2} provides  further evidence for the no-go theorems 
 and is the hallmark of the ``position diffusion'', a known  artifact in the quantum theory of Brownian motion \cite{2009-Vacchini}.

According to Eq.~\eqref{_2-nd_moments_x^2}, $\tder{}{t}\midop{\hat x^2}_{\theta}\big|_{t{=}0}$ is sensitive to smoothness of $\fc_k(\pp)$. Specifically, the rhs of Eq.~\eqref{_2-nd_moments_x^2} is exploded by any highly oscillatory components of $\fc_k(\pp)$ and diverges if $\fc_k(\pp)$ is discontinuous. This entirely quantum effect is the origin of poor performance of the clipped solutions \eqref{__clipping_rule} seen in Fig.~\ref{@FIG.01'}a. Equation~\eqref{_2-nd_moments_x^2} uncovers unavoidable errors in the potential energy. The optimal solutions \eqref{_optimal_f_k} enforce error cancellation $\tder{}{t}\midop{\tfrac{\hat p^2}{2m}}_{\theta}\big|_{t{=}0}{=}{-}\tder{}{t}\midop{U(\hat x)}_{\theta}\big|_{t{=}0}$ between kinetic and potential energies  leaving the total energy intact $\tder{}{t}\midop{\hat H}_{\theta}\big|_{t{=}0}{=}0$. In fact, the error cancellation is achieved with a large class of physically feasible functions $\fc_k(\pp)$ that may substantially differ from the solutions \eqref{_optimal_f_k} everywhere but the region of high probability density $\varpi(p){=}\Tr[\delta{(\hat p{-}p)\rhoth{\theta}}]$ (however, note the remark in \appref{@APP:phys_meaning}). This is illustrated in Fig.~\ref{@FIG.01'}a by dashed curves overlapping with solid curves.


The master equations \eqref{Quantum_Liouville_equation_nD} and \eqref{theorem:trans_inv_Lbd} provide accurate non-perturbative description of collisions with a background gas of atoms or photons \cite{1996-Poyatos,2001-Vacchini-a,2005-Vacchini,2008-Hornberger,
2015-Barchielli}
. Hence, the above theoretical conclusions can be directly tested in the laboratory using well-developed techniques, e.g., the setup shown in Fig.~\ref{@FIG.01'}b. Here a two-level atom is subject to two orthogonally polarized counterpropagating monochromatic nonsaturating laser fields of the same amplitude $\cal E$ and frequency $\omlas$. We show in \appref{@APP:phys_meaning}~that the translational motion of the atom can be modeled using Eq.~\eqref{Quantum_Liouville_equation_nD} with an isotropic friction term of form $\Lrel{=}\Bdn_{\kappa,\fcisoD}$. Here
\begin{gather}\label{_coherent_Doppler_f(p)}
\kappa{=}\tfrac{\omlas}c,~~\fcisoD(p){=}\tilde c_{1}({\tilde c_{2}^2{+}(p{-}\tilde c_{3})^2})^{{-}\frac12},~~\tilde c_{k}{\in}\mathbb{R}
\end{gather}
and the parameters $\tilde c_{k}$ can be tuned by $\cal E$ and $\omlas$. 

Now we are ready to clarify why the deviations from  canonical equilibrium increase with $|\kappa|$ in Fig.~\ref{@FIG.01'}a. The parameters $\hbar|\kappa|$ and $\fcisoD(p)^2$ in Eq.~\eqref{_coherent_Doppler_f(p)} can be regarded as the change of atomic momentum after absorption of a photon and the absorption rate. The case of small $\hbar|\kappa|{\ll}\sqrt{\midop{\hat p^2}}$ and large $\fcisoD(p)^2$ implies tiny and frequent momentum exchanges subject to the central limit theorem. The net result is a velocity-dependent radiation pressure with vanishing fluctuations. The opposite case of large $\hbar|\kappa|{\gg}\sqrt{\midop{\hat p^2}}$ and small $\fcisoD(p)^2$ is the strong shot noise limit, where the stochastic character of light absorption is no longer averaged out, notably perturbing the thermal state. Note that a similar interpretation applies to quantum statistical forces in Ref.~\cite{2016-Vuglar}.

The dissipative model \eqref{Quantum_Liouville_equation_nD} and \eqref{theorem:trans_inv_Lbd} with optimized parameters \eqref{_optimal_f_k} is further analyzed in Fig.~\ref{@FIG.01b'} using the same parameters as in Fig.~\ref{@FIG.01'}a. Both Figs.~\ref{@FIG.01b'} and \ref{@FIG.01'}a indicate that thermalization can be modeled for a wide range of recoil momenta $\hbar\kappa\in\left({-}(\hbar\sqrt{\beta})^{-1},(\hbar\sqrt{\beta})^{-1}\right)$ and the higher the temperature, the better the accuracy. Thus, Eqs.~\eqref{_Bondarian_gen} and \eqref{_Diffusion_operator} enable to simulate a variety of velocity dependences of friction and diffusion. 

Finally, Fig.~\ref{@FIG.01b'}a benchmarks such simulations against the commonly used quantum optical master equation (QOME) \cite{BOOK-Gardiner} defined by Eq.~\eqref{_Lindbladian_definition} with $K{=}2$, $\hat L_1{=}\sqrt{2\Gamma\omega}({1{-}e^{{-}\frac{\hbar\omega}{\theta}}})^{{-}\frac12}\hat a$, $\hat L_2{=}\sqrt{2\Gamma\omega}({e^{\frac{\hbar\omega}{\theta}}}{-}1)^{{-}\frac12}\hat a^{\dagger}$, where $\hat a$ is the harmonic oscillator annihilation operator. For a correct comparison, the parameters of both models are adjusted to ensure identical decay rates in Eq.~\eqref{_stability_check}. Systematic errors in our model and QOME are comparable for the equilibrium displacements $\Delta x_0{\sim}\hbar\beta^{{-}\frac12}$ at zero temperature and $\Delta x_0{\sim}0.1\hbar\beta^{{-}\frac12}$ for $\theta{\sim}\hbar\omega$. For low-frequency molecular vibrational modes ($m{\sim}10^4$\,atomic units, $\omega{\sim}200$\,cm$^{-1}$), these shifts are of order $0.4$\,\AA~and $0.04$\,\AA, respectively, which are in the range of typical molecular geometry changes due to optical excitations or liquid environments. We found the displacement-independent errors in the model \eqref{Quantum_Liouville_equation_nD} and \eqref{theorem:trans_inv_Lbd} 
to be very important for quantum control via reservoir engineering.
Furthermore, the same feature can also be exploited for engineering the mechanical analogs of nonreciprocal optical couplings \cite{2015-Metelmann} and energy-efficient molecular quantum heat machines \cite{2016-Zhdanov}. These subjects will be explored in a forthcoming publication.


\let\section\oldsection

\begin{acknowledgments}
We thank to Alexander Eisfeld for valuable discussions and drawing our attention to Refs.~\cite{1995-Holevo,1996-Holevo,2005-Petruccione,2005-Vacchini,2009-Vacchini}. We are grateful to the anonymous referees for important suggestions and insightful critic. T.~S{.} and D.~V.~Zh{.} thank the National Science Foundation (Award number CHEM-1012207 to T. S.) for support.
D.~I.~B{.} is supported by AFOSR Young Investigator Research Program (FA9550-16-1-0254).
\end{acknowledgments}

\onecolumngrid
\bibsection
\begin{center}
\Large\protect{\texttt{\uppercase{Supplemental material}}}
\end{center}
\twocolumngrid
\appendix
\setcounter{section}{0}
\section*{Introduction}
This supplemental material is organized as follows. In the Sections~\ref{@APP:theo:trans_inv_Lbd}, \ref{@APP:theo:no_go(T=0)} and \ref{@APP:theo:no_go(T>0)} we give the proofs of Lemma~\ref{@theo:trans_inv_Lbd}, first, and second no-go theorems, respectively. Finally, the supporting mathematical derivations for the Doppler cooling model briefly discussed in the main text are provided in Section~\ref{@APP:phys_meaning}.

The Roman numbers in parentheses refer everywhere to the equations in the main text of the letter.

\section{The proof of lemma~\ref{@theo:trans_inv_Lbd}\label{@APP:theo:trans_inv_Lbd}}
\newcommand{\SpSh}{{\cal R}}
The property of the translational invariance can be formulated as
\begin{gather}\label{Quantum_Liouville_trans_inv_nD(reformulated)}
\forall \dxx: {\cal R}_{\dxx}\Lrel{\cal R}_{-\dxx}{=}\Lrel,
\end{gather}
where
\begin{gather} \label{__spatial_shift_superoperator}
\SpSh_{\dxx}{=}e^{{-}\frac{i}{\hbar}\dxx\hat\pp}\odot e^{\frac{i}{\hbar}\dxx\hat\pp}
\end{gather} 
is the superoperator of translational shift: $\forall g(\hat\xx): \SpSh^{\intercal}_{\dxx}[g(\hat\xx)]{=}g(\hat\xx{+}\dxx)$.

With the help of the canonical commutation relations, any operator $\hat L_k{=}L_k(\hat\pp,\hat\xx)$ can expanded in the series
$\hat L_k{=}\sum_{l,m}c_{k,l,m}\hat B_{l,m}$, where $\hat B_{l,m}{=}e^{{-}i\set{\kappa}_l\hat\xx}g_m(\hat\pp)$ and the functions $g_m(\pp)$ constitute a set of (not necessarily orthogonal) basis functions. Using this expansion, any superoperator of form $\Lrel{=}\sum_k\Lbd_{\hat L_k}$ can be rewritten as
\begin{gather}
\Lrel{=}\sum_{k,l_1,m_1,l_2,m_2}c_{k,l_1,m_1}c_{k,l_2,m_2}^*\tilde{\cal L}^{\idx{lbd}}_{\hat B_{l_1,m_1},\hat B_{l_2,m_2}},
\end{gather}
where
\begin{gather}
\tilde{\cal L}^{\idx{lbd}}_{\hat A_1,\hat A_2}{\defeq}\hat A_1\odot\hat A_2^{\dagger}-\frac12(\hat A_1\hat A_2^{\dagger}\odot{+}\odot\hat A_1\hat A_2^{\dagger}).
\end{gather}
It follows from Eq.~\eqref{Quantum_Liouville_trans_inv_nD(reformulated)} that if $\Lrel$ is translationally invariant then it should satisfy the identity
\begin{gather}
\Lrel{=}\left.\frac{1}{(2L)^N}\int_{-L}^{L}...\int_{-L}^{L}\,{\cal R}_{\dxx}\Lrel{\cal R}_{-\dxx}\mathrm{d}^N\diff\dxx\right|_{L{\to}\infty}{=}\notag\\
\sum_{l}\sum_{m_1,m_2}\tilde c^{(l)}_{m_1,m_2}\tilde{\cal L}^{\idx{lbd}}_{\hat B_{l,m_1},\hat B_{l,m_2}},\label{Quantum_Liouville_trans_inv_nD_a}
\end{gather}
where the Hermitian matrices $\tilde c^{(l)}$ are defined as
\begin{gather}
\tilde c^{(l)}_{m_1,m_2}{=}\sum_{k}c_{k,l,m_1}c_{k,l,m_2}^*.
\end{gather}
Let us substitute in Eq.~\eqref{Quantum_Liouville_trans_inv_nD_a} the matrices $\tilde c^{(l)}$ with their Jordan decomposition $\tilde c^{(l)}{=}\tilde u^{(l)}\tilde\gamma^{(l)}{\tilde u^{(l)}}^{\dagger}$, where $\tilde u^{(l)}$ is unitary and $\tilde\gamma^{(l)}$ is diagonal. The result is
\begin{gather}\label{Quantum_Liouville_trans_inv_nD_b}
\Lrel{=}\sum_{l,m}{\cal L}^{\idx{lbd}}_{\hat A_{l,m}},
\end{gather}
where $A_{l,m}{=}\tilde f_{l,m}(\hat\pp)e^{{-}i\set{\kappa}_l\hat\xx}$ and $\tilde f_{l,m}(\hat\pp){=}\sqrt{\tilde\gamma^{(l)}_{m,m}}\times\sum_{m'}\tilde u^{(l)}_{m',m}g_{m'}(\hat\pp)$. Finally, note that Eq.~\eqref{Quantum_Liouville_trans_inv_nD_b} can be cast into the form \eqref{theorem:trans_inv_Lbd} by replacing the compound index $\{l,m\}$ with the single consecutive index $k$. The lemma is proven.
\begin{remark}[Remark 1.]
In this work, the Gaussian (continuous) translationally invariant dissipators of form 
\begin{gather}\label{_Lbd_Gaussian}
{\LvnG}{=}\sum_k\Lbd_{\ehat\AG_k},~~\ehat\AG_k{=}\kkappaG_k\hat\xx{+}\ffG_k(\hat\pp)~~(\kkappaG_k{\in}\mathbb{R}^N)
\end{gather}
are treated as the limiting case of Eq.~\eqref{theorem:trans_inv_Lbd} with $\kkappa_k{=}\epsilon\kkappaG_k{\to}0$. Specifically, one can verify by direct calculation that
\begin{gather}\label{_Lbd_Gaussian[lemma-form]}
\Lbd_{\ehat\AG_k}{=}\left.\Lbd_{\ehat\AG_{k,+}}{+}\Lbd_{\ehat\AG_{k,-}}\right|_{\epsilon{\to}0}{-}i\frac{\hbar}2\kkappaG_k\left[\pder{\ffG_k(\hat\pp)}{\hat\pp},\odot\right],\\
\ehat\AG_{k,\pm}{=}\tfrac1{\sqrt 2}\left(\tfrac{i}{\epsilon}{\pm}\ffG_k(\hat\pp)\right)e^{{\mp}i\epsilon\kkappaG_k\hat\xx}.
\end{gather}
\end{remark}
\begin{remark}[Remark 2.]
The translation invariance criterion is generalized to non-Markovian dynamics in Ref.~\cite{2017-Gasbarri}.
\end{remark}

\section{The proof of no-go theorem~\ref{@theo:no_go(T=0)} (by contradiction)\label{@APP:theo:no_go(T=0)}}
Suppose that some eigenstate $\ket{\Psi_{0}}$ of Hamiltonian $\hat H$ is also the fixed point of the quantum Liouvillian $\cal L$ defined by Eqs.~\eqref{Quantum_Liouville_equation_nD} and \eqref{theorem:trans_inv_Lbd}. Since $\Lrel$ is assumed translation invariant, it should commute with spatial shift superoperator \eqref{__spatial_shift_superoperator} for any $\dxx$. Hence, $\Lrel[\SpSh_{\dxx}[\hat\rho_0]]=\SpSh_{\dxx}[\Lrel[\hat\rho_0]]{=}0$, where $\hat\rho_0{=}\proj{\Psi_{0}}$. Furthermore, the linearity of $\Lrel$ implies that
\begin{gather}\label{__pre_w_g}
\forall g(\xx'): \Lrel[\int g(\xx')\SpSh_{\dxx}[\hat\rho_0]d^N \xx']{=}0.
\end{gather}  
Equation \eqref{__pre_w_g} can be further simplified using the identity
\begin{gather}\label{__state_shift_identity}
e^{{-}\frac{i}{\hbar}\xx'\hat\pp}\ket{\Psi_{0}}{=}\sqrt{2\pi\hbar}\Psi_{0}(\hat\pp)\ket{\xx'},
\end{gather}
where $\ket{\xx'}$ is the eigenstate of position operator: $\hat x_k\ket{\xx'}{=}x_k'\ket{\xx'}$, $\scpr{\xx''}{\xx'}{=}\delta(\xx''{-}\xx')$. 
The validity of Eq.~\eqref{__state_shift_identity} can be verified by comparing the wavefunctions in momentum representation corresponding to its left and right sides. Identities \eqref{__state_shift_identity} and $\int g(\xx')\proj{\xx'}d^N\xx'{=}g(\hat\xx)$ allow to equivalently rewrite Eq.~\eqref{__pre_w_g} as
%
%
%
\begin{gather}\label{_w_g}
\forall g(\xx'): \Lrel[\hat w_g]{=}0,~~~\hat w_g{=}\Psi_{0}(\hat\pp)g(\hat\xx)\Psi_{0}(\hat\pp)^{\dagger}.
\end{gather}
Consider the case $g(\xx){=}g_{\llambda}(\xx){=}e^{-i\llambda\xx}$, where $\llambda$ is some real $N$-dimensional vector. Note that the operator $\Lrel[\hat w_g]$ then includes the explicit dependence on coordinate operators $\hat\xx$ only in forms of matrix exponentials $e^{-i\llambda\hat\xx}$, $e^{\pm i\kkappa_k\hat\xx}$ and commutators $[\hat x_k,\odot]$. Using the commutation relation $e^{-i\tilde\llambda\hat\xx}\hat\pp{=}(\hat\pp{+}\hbar\tilde\lambda)e^{-i\tilde\llambda\hat\xx}$ with $\tilde\llambda{=}\llambda,\pm\kkappa_k$, it is possible to group out the momentum and coordinate operators in $\Lrel[\hat w_g]$ and rewrite the condition \eqref{_w_g} as:
\begin{gather}
0{=}\Lrel[\hat w_{g_{\llambda}}]{=}
\tilde G_{\llambda}(\hat\pp)e^{-i \llambda\hat{\xx}}\label{_Lrel(w_g)},
\end{gather}
where
\begin{gather}\label{__G_lambda}
\tilde G_{\llambda}(\pp){=}G(\pp,\pp{+}\hbar\llambda)\Psi_0 (\pp)\Psi_0(\pp{+}\hbar\llambda)^{*}
\end{gather}
and
\begin{gather}
G(\pp,\pp'){=}\sum_k\left(
F_k(\pp)F_k(\pp')^{*}{-}\tfrac{|\tilde f_k(\pp)|^2{+}|\tilde f_k(\pp')|^2}2
\right){+}\\\notag
\hbar\kkappaaux(\tpder{\ln(\Psi_0 (\pp))}{\pp}{+}\tpder{\ln(\Psi_0 (\pp')^{*})}{\pp'}){-}i(\faux(\pp){-}\faux(\pp')),
\\
F_k(\pp){=}\tilde f_k (\pp{+}\hbar\kkappa_k) \frac{\Psi_0(\pp{+}\hbar\kkappa_k)}{\Psi_0(\pp) }.
\end{gather}
Note that the cancellation of all the matrix exponentials $e^{\pm i\kkappa_k\hat\xx}$ at the rhs of Eq~\eqref{_Lrel(w_g)} is the consequence of the translation invariance of $\Lrel$. Condition \eqref{_Lrel(w_g)} implies that 
\begin{gather}\label{condition_in_Psi_0}
\forall \pp{\in}\mathbb{R}^N,\forall\lambda\in\mathbb{R}: \tilde G_{\lambda}(\pp){=}0,
\end{gather}
and $\forall \pp,\pp'{\in}\mathbb{R}^N: G(\pp,\pp'){=}0$ (except a possible zero measure subset of points $\{\pp,\pp'\}$ where $\Psi_0(\hat\pp)\Psi_0(\hat\pp')^{\dagger}{=}0$). In particular, this means that 
\begin{gather}\label{dG/dp_1dp_2}
\begin{split}
\forall n,\forall \pp,\pp'{\in}\mathbb{R}^N:& \pder{^2}{p_{n}\partial p'_{n}} G(\pp,\pp'){=}\\
&\sum_k
\tpder{F_k^{(n)}(\pp)}{p_n}\left(\tpder{F_k^{(n)}(\pp')}{p_n'}\right)^{*}{=}0.
\end{split}
\end{gather} 
Equality \eqref{dG/dp_1dp_2} can be satisfied only if $\forall k: F_k(\pp){\propto}$const, i.e., if $\tilde f_k (\pp){=}c_k\frac{\Psi_0(\pp{-}\hbar\kkappa_k)}{\Psi_0(\pp)}$, where $c_k$ is some real constant. Substitution of this expression and $\llambda{=}\zzero$ into Eq.~\eqref{__G_lambda} gives $\tilde G_{\zzero}(\pp){=}\sum_kc_k^2\left({|{\Psi_0(\pp)}|^2{-}|\Psi_0(\pp{-}\hbar\kkappa_k)}|^2\right){+}\hbar\kkappaaux\tpder{|{\Psi_0}(\pp)|^2}{\pp}$
\footnote{Using \eqref{_Lbd_Gaussian[lemma-form]}, it is straightforward to deduce that the corresponding summand for Gaussian dissipator \eqref{_Lbd_Gaussian} takes form $\frac12\hbar^2\kkappaG\pder{}{\pp}(\kkappaG\pder{}{\pp}|{\Psi_0(\pp)}|^2)$. The resulting contribution in the lhs of Eq.~\eqref{condition_in_Psi_0_integral} is $|\kkappaG|^2\hbar^2$.
}.
Multiplication of the both sides of Eq.~\eqref{condition_in_Psi_0} by $\pp^2$ and subsequent integration over $\pp$ gives:
\begin{align}\label{condition_in_Psi_0_integral}
\int&\pp^2\tilde G_{\zzero}(\pp)\diff^N\pp{=}
\notag\\&\sum_kc_k^2\hbar^2\kkappa_k^2{-}
2\hbar(\kkappaaux{+}\sum_kc_k^2\kkappa_k)\matel{\Psi_0}{\hat\pp}{\Psi_0}{=}0.
\end{align}
According to our assumption, $\matel{\Psi_0}{\hat\pp}{\Psi_0}{=}\zzero$\footnote{The equality $\matel{\Psi_0}{\hat\pp}{\Psi_0}{=}\zzero$ holds for any non-degenerate eigenstate of the time-reversal invariant Hamiltonian \eqref{Quantum_Hamiltonian_nD}}. 
 Hence, Eq.~\eqref{condition_in_Psi_0_integral} implies that $\sum_kc_k^2|\kkappa_k|^2{=}0$. This equality holds only if $\forall k:\kkappa_k{=}\zzero$. However, in this case all functions $\tilde f_k (\pp){=}c_k$ reduce to constants, so that $\Lrel{=}0$. 
\newcommand{\cchi}{\boldsymbol{\chi}}
This result completes the proof.


\section{The proof of no-go theorem~\ref{@theo:no_go(T>0)} (by contradiction)\label{@APP:theo:no_go(T>0)}}
Denote as $\Psi_{k}(\pp)$ and $E_k$ $(k{=}0,...,\infty)$ the momentum-space wavefunction and energy of the $k$-th eigenstate $\ket{\Psi_{k}}$ of the Hamiltonian $\hat H$. The thermal state $\rhoth{\theta}$ can be expressed in these notations as
\begin{gather}
\rhoth{\theta}{=}\rnorm\sum_ke^{{-}\frac{E_k}{\theta}}\proj{\Psi_{k}}
\end{gather}

Suppose that there exists such relaxation superoperator of form \eqref{theorem:trans_inv_Lbd} that $\Lrel[\rhoth{\theta}]{=}0$.
Owing to assumed linearity and translational invariance of $\Lrel$, the thermal state $\rhoth{\theta}$ should satisfy the relation similar to \eqref{__pre_w_g}:
\begin{gather}\label{__pre_w_(theta,g)}
\forall g(\xx'): \Lrel[\int g(\xx')\SpSh_{\dxx}[\rhoth{\theta}]d^N \xx']{=}0,
\end{gather}
where $\SpSh_{\dxx}$ is the spatial shift superoperator defined by Eq.~\eqref{__spatial_shift_superoperator}. With the help of relation \eqref{__state_shift_identity}, one can apply to Eq.~\eqref{__pre_w_(theta,g)} the same procedure as was used to derive the equality \eqref{_w_g} from Eq.~\eqref{__pre_w_g}. The result is
\begin{gather}\label{_Lrel(_w_{theta,g})=0}
\forall g(x): \Lrel[\hat w_{\theta ,g}]{=}0,
\end{gather}
where
\begin{gather}\label{_w_{theta,g}}
\hat w_{\theta,g}{=}\rnorm\sum_ke^{{-}\frac{E_k}{\theta}}\Psi_{k}(\hat\pp)g(\hat\xx)\Psi_{k}(\hat\pp)^{\dagger}.
\end{gather}

Consider the case $g(\xx){=}g_{\llambda}(\xx){=}e^{{-}i\llambda\xx}$, where $\llambda$ is some real $N$-dimensional vector. The result of application of $\Lrel$ to $\hat w_{\theta,g_{\llambda}}$ can be represented after some algebra as
\begin{gather}\label{__Lrel[w]}
\Lrel[\hat w_{\theta,g_{\llambda}}]{=}G_1\left(\hat\pp{+}\tfrac{\hbar\llambda}{2},\llambda\right)e^{{-}i\llambda\hat\xx},
\end{gather}
where
\begin{align}\label{__G_1(p,lambda)}
G_1(\pp,&\llambda)={-}i\BFn_{\theta}(\pp,\llambda)\left(\faux(\pp{-}\tfrac{\hbar\lambda}2){-}\faux(\pp{+}\tfrac{\hbar\lambda}2)\right){+}\notag\\
&
\hbar\kkappaaux\tpder{\BFn_{\theta}(\pp,\llambda)}{\pp}{+}\sum_k \biggl(Q_{k,n}(\pp{+}\hbar\kkappa_k,\llambda){-}\notag\\
&\tfrac12\BFn_{\theta}(\pp,\llambda)\biggl({\left|\tilde{f}_k\left(\pp{+}\tfrac{\hbar \llambda }{2}\right)\right|^2{+}\left|\tilde{f}_k\left(\pp{-}\tfrac{\hbar \llambda}{2}\right)\right|^2}
\biggr)\biggr)
,
\\
Q_k(\pp,&\llambda )=\BFn_{\theta}(\pp,\llambda)\tilde{f}_k(\pp{-}\tfrac{\hbar \llambda }{2}) \tilde{f}_k^*(\pp{+}\tfrac{\hbar \llambda }{2}).
\end{align}
In derivation of \eqref{__G_1(p,lambda)} the identity
\begin{gather}
\BFn(\pp,\llambda ){=}
\rnorm\sum_k e^{{-}\frac{E_k}{\theta}}\Psi_{k}(\pp{-}\tfrac{\hbar\llambda}{2})\Psi_{k}^{*}(\pp{+}\tfrac{\hbar\llambda}{2})
\end{gather}
was used which follows directly from the definition \eqref{_Blokhintsev_function} of the Blokhintsev function.

Eqs.~\eqref{_Lrel(_w_{theta,g})=0} and \eqref{__Lrel[w]} require that 
\begin{gather}\label{__G_1(p,lambda){=}0}
\forall\pp,\llambda:G_1(\pp,\llambda){=}0,
\end{gather} 
and hence $\forall \llambda: \bar G_2(\llambda){=}\int_{{-}\infty}^{\infty}\ldots\int_{{-}\infty}^{\infty}\diff^N \pp\,G_2(\pp,\llambda){=}0$,
where
\begin{gather}\label{__G_2(p,lambda)}
\begin{split}
G_2(&\pp,\llambda){=}G_1(\pp,\llambda){+}G_1(\pp,{-}\llambda){=}
\\
&\sum_k\biggl\{
{-}\left|\tilde{f}_k\left(\pp{+}\tfrac{\hbar\llambda }{2}\right){-}\tilde{f}_k\left(\pp{-}\tfrac{\hbar \llambda}{2}\right)\right|^2\BFn_{\theta}(\pp,\llambda){+}
\\
&\sum_{\alpha,\beta{=}\pm1}\beta Q_k\left(\pp{+}\tfrac{\beta{+}1}2\hbar\kkappa_k,\alpha\llambda\right)\biggr\}
{+}2\hbar\kkappaaux\tpder{\BFn_{\theta}(\pp,\llambda)}{\pp}
.
\end{split}
\end{gather}

The last equality in \eqref{__G_2(p,lambda)} is obtained assuming that $\BFn_{\theta}(\pp,{-}\llambda){=}\BFn_{\theta}(\pp,\llambda)$ (see Eq.~\eqref{_B(p,lambda)-features-a}).
It is easy to check that the integrations over all terms in the last line of \eqref{__G_2(p,lambda)} cancel out, so that
\begin{align}
\bar G_2(\llambda){=}&{-}\int_{{-}\infty}^{\infty}\ldots\int_{{-}\infty}^{\infty}\diff^N \pp{\times}\notag\\
&\sum_k\left|\tilde{f}_k\left(\pp{+}\tfrac{\hbar\llambda }{2}\right){-}\tilde{f}_k\left(\pp{-}\tfrac{\hbar\llambda}{2}\right)\right|^2\BFn(\pp,\llambda).\label{_bar_G_2(p,lambda)}
\end{align}
According to the assumption \eqref{_B(p,lambda)-features-a}, the integrand in \eqref{_bar_G_2(p,lambda)} is nonnegative. Moreover, $\bar G_2(\llambda){=}0$ iif $\forall k: \tilde f_k(\pp){=}c_k{=}$const. Hence, the expression \eqref{__G_1(p,lambda)} for $G_1(\pp,\llambda)$ can be simplified as
\begin{gather}\label{__G_1_simplified(p,lambda)}
G_1(\pp,\llambda )=\sum _k c_k^2 \left(\BFn\left(\pp{+}\hbar\kkappa_k,\llambda \right){-}\BFn(\pp,\llambda )\right).
\end{gather}
Note that the terms $\Lbd_{\hat A_k}$ in Eq.~\eqref{theorem:trans_inv_Lbd} with $\tilde f_k(\pp){=}$const will have non-trivial effect only if $\kkappa_k{\ne}0$%
\footnote
{In the case of Gaussian dissipator \eqref{_Lbd_Gaussian} Eq.~\eqref{__G_1_simplified(p,lambda)} reduces to
\begin{gather}\label{__G_1_simplified(p,lambda)[Gaussian]}
G_1(\pp,\llambda)=\frac12\hbar^2\sum_{k,m,n}\mu_{k,n}\mu_{k,m}\pder{^2}{p_np_m}\BFn(\pp,\llambda ).\tag{\ref*{__G_1_simplified(p,lambda)}*}
\end{gather}
By assumption \eqref{_B(p,lambda)-features-b}, the quadratic form $\pder{^2}{p_np_m}\BFn(\pp,\llambda )$ in \eqref{__G_1_simplified(p,lambda)[Gaussian]} is negative-definite at $\{\pp,\llambda \}{=}\{\zzero,\zzero\}$. Hence, $G_1(\zzero,\zzero){<}0$, which contradicts Eq.~\eqref{__G_1(p,lambda){=}0} and completes the proof for this case.
}%
. However, it follows from \eqref{_B(p,lambda)-features-b} that in this case $G_1(\zzero,\zzero){<}0$ which contradicts Eq.~\eqref{__G_1(p,lambda){=}0}. The theorem is proven.

\section{Testing the model \texorpdfstring{(\ref{Quantum_Liouville_equation_nD}) and  (\ref{theorem:trans_inv_Lbd})}{} in the laboratory\label{@APP:phys_meaning}}

In this section, we provide the detailed analysis of the Doppler cooling example introduced in the main text (see Fig.~\ref{@FIG.01'}b in the main text) and prove that the cooling mechanism is the quantum friction of form \eqref{theorem:trans_inv_Lbd-isotropic}. 

In the proposed setup an atom is subject to two orthogonally polarized counterpropagating beams of the same field amplitude $\cal E$ and carrier frequency $\omlas$ (hereafter in
this section we will omit the subscript l for shortness since it will not cause any ambiguity). We assume that $\omega$ is close to the frequency $\omega_{\idx{a}}$ of the transition $\es{g}{\LR}\es{e}$ between the ground $\es{g}$ and degenerate excited $\es{e}$ electron states of $s$- and $p$-symmetries, respectively. Let $d$ be the absolute value of the transition dipole moment and $\gamma$ be the excited state spontaneous decay rate. 

\begin{figure}[tbp]
\centering\includegraphics[width=0.7\columnwidth]
{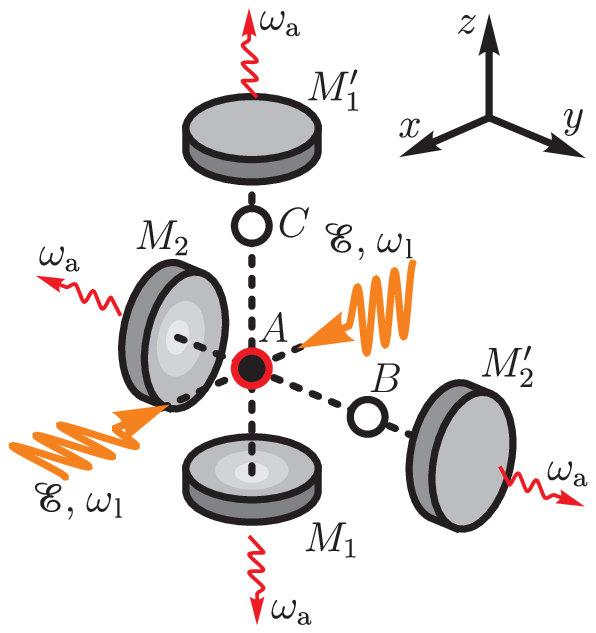}
\caption{The possible Doppler cooling setup where stochastic recoil accompanying the spontaneous emission is damped along the $x$-axis. Here the atom of interest $A$ is put into intersected orthogonal optical cavities formed by pairs of mirrors $M_1$, $M_1'$ and $M_2$, $M_2'$. The cavities are tuned resonant to the atomic $\es{g}\LR\es{e}$ transition and force atom to spontaneously emit absorbed photons predominantly in the directions perpendicular to the $x$-axis via the Purcell effect. The decay rate $\gamma$ can be controlled by changing the cavities Q-factors. The collateral increase of the energy of motions along $y$- and $z$-axes is restricted by sympathetic cooling by two auxiliary atoms $B$ and $C$.\label{@FIG.A01}}
\end{figure}

For the spatial arrangement depicted in Fig.~\ref{@FIG.01'}b the translation motion of the atom along $x$-axis is coupled to the field-induced electron dynamics since each absorbed
or coherently emitted photon changes the $x$-component of atomic momentum hereafter denoted as $p$. Furthermore, we will assume that the spontaneous decay does not affect the $x$-component of atomic momentum. The latter condition can be achieved using, e.g., an arrangement shown in Fig.~\ref{@FIG.A01}.

The master equation which describes this coupled dynamics can be written within the rotating wave approximation in the form \eqref{Quantum_Liouville_equation_nD} with  

\begin{gather}
\begin{split}
\hat H{=}&\frac{\hat p^2}{2m}{-}\hbar\omega_{\idx{a}}\proj{\es{g}}{+}
\biggl\{\xi_1(t)\proj[\es{e}_1]{\es{g}}e^{-i(\omega t{-}\kappa \hat x)}{+}\\&
\xi_2(t)\proj[\es{e}_2]{\es{g}}e^{-i(\omega t{+}\kappa \hat x)}{+}\mbox{h.c.}
\biggr\}
\end{split}
\end{gather}
and
\begin{gather}\label{doppler_L_rel}
\Lrel{=}{\gamma}\sum_{n{=}1}^2\Lbd_{\proj[\es{g}]{\es{e}_n}}.
\end{gather}
Here $\xi_k(t){=}{-}\frac12\vec d_k\vec{\cal E}_k(t)$, where $\vec d_1$ and $\vec d_2$ are the transition dipole moments associated with the $s{\to}p_z$ and $s{\to}p_y$ electronic transitions into degenerate electronically excited sublevels $\es{e}_1$ and $\es{e}_2$, respectively, and $\vec{\cal E}_k(t)$ is the slowly varying complex amplitude of the associated field component. The remaining notations are defined in the main text.

\newcommand{\evsop}[2]{\mathop{{{\cal U}^{#1}_{#2}}}}
\newcommand{\evsopt}[2]{\mathop{{{\cal U}^{#1}_{#2}}^{\intercal}}}
\newcommand{\TOdir}{\stackrel{\Rightarrow}{{\cal T}}}
\newcommand{\TOinv}{\stackrel{\Leftarrow}{{\cal T}}}
\newcommand{\Pg}{\hat P_{\es{g}}}
The mean value of any observable of form $\hat O{=}f(\hat p,\hat x)$ can be written in Heisenberg representation as:
\begin{gather}\label{_<O>}
\midop{\hat O(t)}{=}\Tr[\hat\rho_0\evsopt{\cal L}{t,t_0}[\hat O]],
\end{gather}
where we define:
\begin{gather}\label{_evsop}
\forall {\cal L}(t): \evsop{\cal L}{t,t_0}\stackrel{\idx{def}}{=}\TOdir e^{\int_{t{=}t_0}^{t}{\cal L}\diff t}.
\end{gather} 
The symbol $\TOdir$ in \eqref{_evsop} denotes the chronological ordering superoperator which arranges operators in direct (inverse) time order for $t{>}t_0$ ($t{<}t_0$). Let us also define the following notations for the interaction representation generated by arbitrary splitting ${\cal L}(t){=}{\cal L}_0+{\cal L}_1(t)$:
\begin{gather}\label{superoperator_interaction_representation}
(\evsop{\cal L}{t,0})^{\intercal}{=}\evsop{(\cal L_0^{\intercal})}{t,0}\evsop{({\cal L}_{\idx{I}}^{\intercal})}{t,0},
\end{gather} 
where the interaction Liouvillian reads
\begin{gather}\label{interaction_Liouvillian}
{\cal L}_{\idx{I}}^{\intercal}(\tau){=}{\evsop{({\cal L}^{\intercal}_0)}{{-}\tau,0}}{\cal L}_1^{\intercal}(t{-}\tau){\evsop{({\cal L}_0^{\intercal})}{\tau,0}}.
\end{gather}
In the case ${\cal L}_0'{=}\frac{-i}{\hbar}[\frac{\hat p^2}{2m}{-}\hbar\omega_{\idx{a}}\proj{\es{g}},\odot]$ the associated interaction liouvillian \eqref{interaction_Liouvillian} in the rotating wave approximation takes the form:
\begin{gather}\label{L_I'-}
{\cal L}_{\idx{I}}'{\simeq}\frac{-i}{\hbar}[\hat H',\odot]{+}\sum_{n{=}1}^2\Lbd_{\proj[\es{g}]{\es{e}_n}},
\end{gather}
where
\begin{gather}
\hat H'(\tau){=}\sum_{n{=}1}^2\hat\chi_n(\tau)\proj[\es{g}]{\es{e}_n}{+}\mbox{h.c.};\\
\hat\chi_1(\tau){=}\xi_1^*(t{-}\tau)e^{i(\omega t{-}\kappa \hat x{-}(\Delta{-}\frac{\kappa \hat{p}}{m})\tau) };\\
\hat\chi_2(\tau){=}\xi_2^*(t{-}\tau)e^{i(\omega t{+}\kappa \hat x{-}(\Delta+\frac{\kappa \hat{p}}{m})\tau)},
\end{gather}
and $\Delta{=}\omega{-}\omega_{\idx{a}}$ is detuning of carrier frequency of radiation from atomic resonance in the case of system at rest.
Repeated application of the transformation \eqref{superoperator_interaction_representation} to \eqref{L_I'-} with ${\cal L}_0''{=}\Lrel{=}\gamma\sum_{n{=}1}^2\Lbd_{\proj[\es{g}]{\es{e}_n}}$ leads to expression:
\begin{gather}
(\evsop{\cal L}{t,0})^{\intercal}{=}\evsop{{\cal L_0'}^{\intercal}{+}\Lrel^{\intercal}}{t,0}\evsop{({{\cal L}_{\idx{I}}''}^{\intercal})}{t,0},
\end{gather}
so that 
\begin{gather}
\midop{\hat O(t)}{=}\Tr[(\evsop{{\cal L_0'}{+}\Lrel}{t,0}[\hat\rho_0])\evsop{({{\cal L}_{\idx{I}}''}^{\intercal})}{t,0}[\hat O]]\stackrel{t{\gg}\gamma^{-1}}{=}\\
\Tr[\Pg{(\evsop{{\cal L_0'}{+}\Lrel}{t,0}[\hat\rho_0])}\Pg({\evsop{({{\cal L}_{\idx{I}}''}^{\intercal})}{t,0}[\hat O]})\Pg]\label{<O(t)>-doppler(t->inf)},
\end{gather}
where $\Pg{=}\proj{\es{g}}$ and the last equality is due to the exponential damping of excited states populations induced by  relaxation superoperator \eqref{doppler_L_rel}.
Let us consider the evolution $\hat O(t)$ generated by the superoperator $\evsop{{{\cal L}_{\idx{I}}''}^{\intercal}}{t+\delta t,t}$:
\begin{gather}\label{generator_2-order-expansion}
\begin{split}
\hat O(t{+}&\delta t){\simeq}\biggl(1{+}\int_t^{t{+}\delta t}{{\cal L}_{\idx{I}}''}^{\intercal}(\tau)\diff\tau{+}\\
&\int_t^{t{+}\delta t}\diff\tau_2\int_t^{\tau_2}d\tau_1{{\cal L}_{\idx{I}}''}^{\intercal}(\tau_2){{\cal L}_{\idx{I}}''}^{\intercal}(\tau_1)\biggr)\hat O(t).
\end{split}
\end{gather}
Integrands in Eq.~\eqref{generator_2-order-expansion} include the terms oscillating at frequencies $|\Delta{\pm}\frac{k \midop{\hat{p}}}m|$. In sequel we will consider the so-called weak-field limit when these oscillations are rapid relative to the characteristic timescales of the relevant processes, so that the contributions of the associated terms asymptotically vanish. In this limit, the second term in rhs of Eq.~\eqref{generator_2-order-expansion} disappears. The remaining terms constitute two decoupled evolution equations for the reduced density matrices $f_{\es{x}}(\hat p,\hat x,t{+}\delta t){=}\matel{\es{x}}{\hat O(t)}{\es{x}}$ ($\es{x}{=}\es{g},\es{e}$):
\begin{gather}\label{f_g(t)-func}
\begin{split}
f_{\es{g}}(\hat p,&\hat x,t{+}\delta t){=}\biggl({\odot}{+}\frac{1}{\hbar^2}\int_t^{t{+}\delta t}\diff\tau_2\int_t^{\tau_2}d\tau_1e^{\frac{1}{2} \gamma  (\tau_1{-}\tau_2)}\times\\
&\sum_{n{=}1}^2\biggl\{\hat\chi_n(\tau_2){\odot}{\hat\chi_n^{\dagger}(\tau_1)}{+}{\hat\chi_n(\tau_1)}{\odot}\hat\chi_n^{\dagger}(\tau_2){-}\\
&{\odot}{\hat\chi_n(\tau_1)}{\hat\chi_n^{\dagger}(\tau_2)}{-}\hat\chi_n(\tau_2) {\hat\chi_n^{\dagger}(\tau_1)}{\odot}\biggr\}\biggr)[f_{\es{g}}(\hat p,\hat x,t)];
\end{split}\\
f_{\es{e}}(\hat p,\hat x,t{+}\delta t){=}{\cal G}[f_{\es{e}}(\hat p,\hat x,t)]
\end{gather}
The explicit form of $\cal G$ is irrelevant in view of Eq.~\eqref{<O(t)>-doppler(t->inf)}. The first two terms in the curly brackets in Eq.~\eqref{f_g(t)-func} can be transformed as
\begin{subequations}\label{f_g(t)-term}
\begin{gather}
\begin{split}\label{f_g(t)-1-st_term}
\hat\chi_1&(\tau_2){f_{\es{g}}(\hat p,\hat x,t)}{\hat\chi_1^{\dagger}(\tau_1)}{=}\\
&\xi_1^*(t{-}\tau_2)\xi_1(t{-}\tau_1)f_{\es{g}}(\hat{p}{+}\hbar\kappa,\hat{x}{+}\tfrac{\hbar\kappa}{m}\tau_2,t)e^{i \Delta_1(\hat p)(\tau_1{-}\tau_2)}{=}\\
&\xi_1^*(t{-}\tau_2)\xi_1(t{-}\tau_1)e^{i\hat\Delta_1(\hat p)(\tau_1{-}\tau_2) }f_{\es{g}}(\hat{p}{+}\hbar\kappa,\hat{x}{+}\tfrac{\hbar\kappa}{m}\tau_1,t),
\end{split}
\end{gather}
\begin{gather}
\begin{split}\label{f_g(t)-2-nd_term}
\hat\chi_1&(\tau_1){f_{\es{g}}(\hat p,\hat x,t)}{\hat\chi_1^{\dagger}(\tau_2)}{=}\\
&\xi_1(t{-}\tau_2)\xi_1^*(t{-}\tau_1)f_{\es{g}}(\hat{p}{+}\hbar\kappa,\hat{x}{+}\tfrac{\hbar\kappa}{m}\tau_1,t)e^{{-}i \Delta_1(\hat p)(\tau_1{-}\tau_2)}{=}\\
&\xi_1(t{-}\tau_2)\xi_1^*(t{-}\tau_1)e^{{-}i\hat\Delta_1(\hat p)(\tau_1{-}\tau_2) }f_{\es{g}}(\hat{p}{+}\hbar\kappa,\hat{x}{+}\tfrac{\hbar\kappa}{m}\tau_2,t),
\end{split}
\end{gather}
\end{subequations}
where
$
\Delta_1(p){=}\Delta{-}\frac{\kappa( p{+}\frac{\hbar\kappa}{2})}{m}.
$
The extra displacements $\frac{\hbar\kappa}{m}\tau_n$ in the $x$-dependencies of $f_{\es{g}}$ in Eqs.~\eqref{f_g(t)-term} account for the change of the velocity of atom after the photon absorption.
These displacements are typically very small compared to the characteristic scales of spatial change of the function $f_{\es{g}}$ and can be neglected. With this approximation, the exponentials and functions $f_{\es{g}}$ in Eqs.~\eqref{f_g(t)-term} commute, which allows to write:
\begin{subequations}\label{approximation_for_f_g(t)-term}
\begin{gather}
\begin{split}
\frac{1}{\hbar^2}\int_t^{t{+}\delta t}&\diff\tau_2\int_t^{\tau_2}d\tau_1e^{\frac{1}{2}\gamma(\tau_1{-}\tau_2)}{\times}\\
&\left(\hat\chi_1(\tau_2){\odot}{\hat\chi_1^{\dagger}(\tau_1)}{+}{\hat\chi_1(\tau_1)}{\odot}\hat\chi_1^{\dagger}(\tau_2)\right)[f_{\es{g}}(\hat{p},\hat{x},t)]{\simeq}\\
&2C_{+}(\hat p,t)f_{\es{g}}(\hat{p}{+}\hbar\kappa,\hat{x},t)C_{+}(\hat p,t)\delta t,
\end{split}
\end{gather}
\begin{gather}
\begin{split}
\frac{1}{\hbar^2}\int_t^{t{+}\delta t}&\diff\tau_2\int_t^{\tau_2}d\tau_1e^{\frac{1}{2}\gamma(\tau_1{-}\tau_2)}{\times}\\
&\left(\hat\chi_2(\tau_2){\odot}{\hat\chi_2^{\dagger}(\tau_1)}{+}{\hat\chi_2(\tau_1)}{\odot}\hat\chi_2^{\dagger}(\tau_2)\right)[f_{\es{g}}(\hat{p},\hat{x},t)]{\simeq}\\
&2C_{-}(\hat p,t)f_{\es{g}}(\hat{p}{-}\hbar\kappa,\hat{x},t)C_{-}(\hat p,t)\delta t,
\end{split}
\end{gather}
\end{subequations}
where
\begin{subequations}\label{formulas_for_C+-}
\begin{widetext}
\begin{align}\label{-C+-}
C_{+}(p,t)&{=}
\sqrt{s_+(p){+}s_+^{*}(p)},   &s_+(p)&{=}\frac{1}{2\hbar^2\delta t}\int_t^{t{+}\delta t}\diff\tau_2\int_t^{\tau_2}d\tau_1\xi_1^*(t{-}\tau_2)\xi_1(t{-}\tau_1)e^{(i\Delta_1(p){+}\frac{\gamma}2)(\tau_1{-}\tau_2)},
\\
C_{-}(p,t)&{=}\sqrt{s_-(p){+}s_-^{*}(p)},   & s_{-}(p)&{=}\frac{1}{2\hbar^2\delta t}\int_t^{t{+}\delta t}\diff\tau_2\int_t^{\tau_2}d\tau_1\xi_2^*(t{-}\tau_2)\xi_2(t{-}\tau_1)e^{(i\Delta_1({-}p){+}\frac{\gamma}2)(\tau_1{-}\tau_2)}.
\end{align}
\end{widetext}
\end{subequations}
Substitution of approximations \eqref{approximation_for_f_g(t)-term} into \eqref{f_g(t)-func} gives:
\begin{gather}\label{effective_Liouvillian}
f_{\es{g}}(\hat p,\hat x,t{+}\delta t){=}\evsop{{\cal L}_{\idx{eff}}^{\intercal}}{t{+}\delta t,t}[f_{\es{g}}(\hat p,\hat x,t)],
\end{gather}
where
\begin{gather}
{\cal L}_{\idx{eff}}(t){=}{-}\frac{i}{\hbar}[\hat H_{\idx{eff}},\odot]{+}\Lrel^{\idx{eff}},
\end{gather}
\begin{gather}
\label{effective_friction}
\Lrel^{\idx{eff}}{=}
\Lbd_{e^{i\kappa \hat x}C_{+}(\hat p,t)}{}+\Lbd_{e^{{-}i\kappa \hat x}C_{-}(\hat p,t)},
\end{gather}
\begin{gather}
\hat H_{\idx{eff}}{=}i\hbar\sum_{m=\pm}(s_m(\hat p)-s_m^{*}(\hat p)).
\end{gather}
Eq.~\eqref{effective_Liouvillian} allows to calculate the averaging in \eqref{<O(t)>-doppler(t->inf)} within the reduced Hilbert space which involves only the translational degree of freedom:
\begin{gather}
\midop{\hat O(t)}\stackrel{t{\gg}\gamma^{-1}}{=}\Tr[\hat\rho_0^{\idx{red}}{\evsop{\frac{i}{\hbar}[\frac{\hat p^2}{2m},\odot]}{t,0}\evsop{{\cal L}_{\idx{eff}}^{\intercal}}{t,0}[\hat O]}]_{\idx{spatial}}\label{<O(t)>-spatial_only}.
\end{gather}
Here $\hat\rho_0^{\idx{red}}{=}\Tr[\hat\rho]_{\idx{el}}$ whereas $\Tr[\odot]_{\idx{el}}$ and $\Tr[\odot]_{\idx{spatial}}$ denote the partial traces over the electronic and translational subsystems.

The dissipator \eqref{effective_friction} reduces to the isotropic friction of form \eqref{theorem:trans_inv_Lbd-isotropic} provided that
\begin{gather}
\forall p:C_{+}({-}p,t){=}C_{-}(p,t){=}\fciso(p).
\end{gather}
It is easy to verify that this condition is realized in two important cases.

\subsection{Weak coherent laser driving}
In this regime, $\xi_1(t){=}\xi_2(t){=}\xi{=}$const, and there exists such $\delta t$ in the range of applicability of the second-order expansion \eqref{generator_2-order-expansion} that $\delta t{\gg}\gamma^{-1}$. Thence, the integrals in \eqref{formulas_for_C+-} can be easily computed, which gives:
\begin{gather}\label{_coherent_C_{+-}}
\Lrel^{\idx{eff}}{=}\Bdn_{\kappa,\fciso},~~
\fciso(p){=}{\frac{|\xi|}{\hbar}\frac{\sqrt{\gamma/2}}{\sqrt{(\frac{\gamma}2)^2{+} \Delta_1^2(-p)}}},\\
\hat H_{\idx{eff}}{=}{-}\frac{|\xi|^2}{\hbar}\sum_{\alpha{=}\pm1}\frac{\Delta_1(\alpha \hat p)}{(\frac{\gamma}2)^2+ \Delta_1^2(\alpha \hat p)}.
%
%
\end{gather}
Note what the Hamiltonian $\hat H_{\idx{eff}}$ describes the effect of the optical quadratic Stark shift which also can induce the effective potential forces on the system in the case of spatially non-uniform fields $\xi{=}\xi(x)$. 

\subsection{Incoherent driving}
Suppose that the the atom is illuminated by the two classical light sources with the equal spectral densities $I(\omega)$ at the atomic site and having coherence times in the range $\Delta_1^{-1}(p){\ll}t_{\idx{coh}}{\ll}\gamma^{-1}$. In this case, $\xi_{1}(t)$ and $\xi_{2}(t)$ represent the uncorrelated stationary stochastic processes. This allows one to choose such $\delta t$, that $\gamma^{-1}{\gg}\delta t{\gg}t_{\idx{coh}}$, and calculate the integrals in Eqs.~\eqref{formulas_for_C+-} neglecting the terms $\frac{\gamma}2$ in the exponents, which gives
\begin{gather}
\Lrel^{\idx{eff}}{=}\Bdn_{\kappa,\fciso},~~
\fciso(p){=}\frac{\pi d}{\hbar}\sqrt{\frac{1}{2 c}I(\omega{+}\Delta_1(-p))},
\end{gather}
where $I(\omega)$ is the spectral density of each beam. Also, here we assumed equal transition dipole momenta: $d{=}|\vec d_1|{=}|\vec d_2|$.

\begin{remark}[Remark.]
The setup sketched in Fig.~\ref{@FIG.A01} as well as in Fig.~\ref{@FIG.01'} of the main text in principle can be used to measure both the momenta and positions of the environmental photons by registering the scattered photons and the position of atom. This implies that there must exist the fundamental restrictions on the physically admissible shapes and smoothness of profiles $\fciso(p)$ and, more generally, on admissible forms of operators $\hat L_k$ in Eq.~\eqref{_Lindbladian_definition}, that would prevent these measurements from violating the Heisenberg uncertainty principle. The detailed analysis of implications of this important observation is way beyond the scope of this paper and will be the subject of future work.
\end{remark}

\bibliography{quantum_friction_2}

\end{document}